\documentclass[]{aastex631}
\usepackage{graphicx}
\usepackage{natbib}
\usepackage{amssymb,amsmath}
\usepackage{savesym}
\savesymbol{tablenum}
\usepackage{siunitx}
\restoresymbol{SIX}{tablenum}
\usepackage{color}

\usepackage{multirow}

\shorttitle{}
\shortauthors{Pallister \& Jeffrey}

\begin{document}

\title{Exploring the Origin of Solar Energetic Electrons I: Constraining the Properties of the Acceleration Region Plasma Environment}

\author[0000-0002-2376-6725]{Ross Pallister}
\affiliation{Department of Mathematics, Physics \& Electrical Engineering, Northumbria University,  \\
Newcastle upon Tyne, UK\\
NE1 8ST}

\author[0000-0001-6583-1989]{Natasha L. S. Jeffrey}
\affiliation{Department of Mathematics, Physics \& Electrical Engineering, Northumbria University,  \\
Newcastle upon Tyne, UK\\
NE1 8ST}

\begin{abstract}
Solar flare electron acceleration is an efficient process, but its properties (mechanism, location) are not well constrained. Via hard X-ray (HXR) emission, we routinely observe energetic electrons at the Sun, and sometimes we detect energetic electrons in interplanetary space. We examine if the plasma properties of an acceleration region (size, temperature, density) can be constrained from in-situ observations, helping to locate the acceleration region in the corona, and infer the relationship between electrons observed in-situ and at the Sun. We model the transport of energetic electrons, accounting for collisional and non-collisional effects, from the corona into the heliosphere (to 1.0 AU). In the corona, electrons are transported through a hot, over-dense region. We test if the properties of this region can be extracted from electron spectra (fluence and peak flux) at different heliospheric locations. We find that cold, dense coronal regions significantly reduce the energy at which we see the peak flux and fluence for distributions measured out to 1.0 AU, the degree of which correlates with the temperature and density of plasma in the region. Where instrument energy resolution is insufficient to differentiate the corresponding peak values, the spectral ratio of [7-10) to [4-7) keV can be more readily identified and demonstrates the same relationship. If flare electrons detected in-situ are produced in, and/or transported through, hot, over-dense regions close to HXR-emitting electrons, then this plasma signature should be present in their lower-energy spectra (1–20 keV), observable at varying heliospheric distances with missions such as Solar Orbiter.
\end{abstract}

\section{Introduction}

The Sun is an efficient particle accelerator capable of producing kilo-, Mega- and Giga-electronvolt (keV, MeV, GeV) energetic particles \citep[e.g., ][]{holman2011implications,2011SSRv..159..167V,2017SSRv..212.1107K}, in transient events such as solar flares and coronal mass ejections (CME) via magnetic reconnection \citep[e.g., ][]{1957JGR....62..509P,1958IAUS....6..123S,2000mare.book.....P} and/or shocks \citep[e.g., ][]{2006SSRv..123..251F}. In solar flares, a large fraction ($\sim 10-50$\%) of the released magnetic energy is converted into energetic particles, including energetic electrons \citep[e.g., ][]{2012ApJ...759...71E}. However, the processes that accelerate and transport energetic electrons are widely debated with the exact configuration(s) of the acceleration mechanism(s), environment and location(s) still undetermined.

In this standard model, hard X-ray (HXR) producing energetic electrons \citep[e.g., ][]{kontar2011deducing} are transported along newly formed magnetic field lines in the corona, precipitating into the dense layers of the lower atmosphere and losing energy. So-called flare-produced solar energetic electrons (SEEs) can also be detected in the heliosphere \citep[e.g., ][]{1985SoPh..100..537L}, either via in-situ measurements or by their radio emissions \citep[c.f., ][]{2008A&ARv..16....1P}, but the connection between these distinct electron populations, and indeed their connecting magnetic topology, is still poorly understood. Moreover, multi-messenger diagnostics, whether remote sensing or in-situ, can be complicated by various particle propagation and emission effects such as e.g., Coulomb collisions \citep[e.g., ][]{2014ApJ...787...86J}, X-ray albedo \citep[e.g., ][]{1978ApJ...219..705B}, radio wave scattering \citep[e.g., ][]{2017NatCo...8.1515K}, turbulence (parallel and cross-field diffusion) \citep[e.g.,][]{2014ApJ...780..176K} and field line meandering \citep[e.g., ][]{2016A&A...591A..18L}. HXR-emitting electrons are produced close to or within hot and dense flaring loops, possibly via turbulence \citep{2017PhRvL.118o5101K,2021ApJ...923...40S}, and we see the signature of that plasma in their low-energy spectra ($\le30$~keV). Non-thermal electrons accelerated out of a hot, dense plasma and/or non-thermal electrons moving through a hot, dense region and undergoing (partial-) thermalization at lower-energies ($\le50$~keV) \citep[e.g., ][]{2019ApJ...880..136J,2015ApJ...809...35K}, will retain or imprint the properties of that plasma environment.

Most studies to date explore the connection between different flare electron populations by studying the properties of the higher energy power law (usually above $40$~keV). After the launch of the Ramaty High Energy Solar Spectroscopic Imager (RHESSI; \citet{2002SoPh..210....3L}), several studies examined flares with both HXR-producing electrons and SEEs detected at 1.0~AU. \citet{2007ApJ...663L.109K} compared HXR spectra with WIND/3DP \citep{1995SSRv...71..125L} electron spectra at $\sim1.0$~AU. For so-called `prompt' events, where the SEE release time appears to coincide with the flare HXR burst, \citet{2007ApJ...663L.109K} found a clear correlation for both power-law spectral indices and total number of electrons, which is consistent with a single process accelerating both electron populations. Under the assumption that both HXR-producing and escaping electron populations are accelerated by a similar mechanism within similar plasma conditions, we expect near identical spectra. However, in \citet{2007ApJ...663L.109K}, the peak flux spectrum was harder then the inferred cold-thick-target electron spectra at the Sun possibility suggesting the effects of coronal or heliospheric transport processes. A near-identical study using flare data from solar cycle 24, \citet{2021A&A...654A..92D}, also found a strong correlation of about 0.8 between remote HXR-producing and in-situ spectral indices.  \citet{2021A&A...654A..92D} observed an increased correlation for events with `significant anisotropy', suggesting that transport effects reduce the signature of the acceleration region properties in the data. In \citet{2021ApJ...913...89W}, sixteen SEE producing flares were examined; they determined that the spectral index of HXR-producing electrons was no less than the observed high-energy spectral index of SEEs (above a spectral break energy), showing a positive correlation with the high-energy spectral index of SEEs. Further, the spectral analysis (extending down to $\sim5$~keV) suggested that the source of SEEs was high in the corona at a heliocentric distance of $\ge1.3$ solar radii. In contrast, other studies looking at Type III radio bursts \citep[e.g., ][]{2011A&A...529A..66R} offer conflicting results, suggesting that outward propagating electrons are accelerated between heights of $40-60$~Mm ($\approx1.07$ solar radii), much closer to the flare location at the Sun. 

Only with a better understanding of the acceleration environment and transport processes therein can we explore the connection between flare-accelerated electron populations. To this end, we aim to constrain the properties of an acceleration region (e.g., its size, temperature, density, turbulence profile); properties not currently constrained well by remote flare observation or in-situ detection alone. In this initial study (Paper I) we use solely SEE transport modelling in the corona and heliosphere alone to investigate the possibility of extracting acceleration region plasma properties from in-situ data, constraining the likely location of their acceleration, and hence relation with HXR-producing electrons. The study mainly concentrates on the electron spectral observations at heliospheric locations of $0.4-1.0$~AU, that can now be explored with both Parker Solar Probe (PSP; \citet{2016SSRv..204....7F}) and Solar Orbiter (SolO; \citet{2020A&A...642A...1M}). Later studies will combine these constraints with those derived from coronal HXR-centric studies to explore the possibility of a unified acceleration region parameter range for prompt flare events.
We examine if more attention should be given to the understudied electron spectral range of $1-30$~keV, and if signatures of hot and dense plasma can be extracted from in-situ datasets. Most studies concentrate on studying and comparing the properties of interplanetary electrons with their HXR-emitting counterparts above $20-30$~keV only. However, the analysis of HXR-producing electrons shows that the lower portion of the spectrum $\le 20$~keV is directly related to the presence and properties of the surrounding coronal plasma \citep[e.g.,][]{2015ApJ...809...35K}.  We suggest that any hot, over-dense plasma signature will help to locate the origin of such particles and in particular their relationship to HXR-emitting electrons. For this task, we assume that electron spectra (peak flux and fluence) in the range of $1-20$~keV are appropriate for this purpose.

In Section \ref{methods} we present the model of electron transport for both the coronal region and heliospheric components, and the coronal plasma properties used in the former. In Section \ref{results} we present the results of the study, including different diagnostics that can be used to estimate the acceleration plasma environment from a partially-thermalized electron distribution. In Section \ref{discussion}, we summarize and discuss the main results. 

\section{Coronal and heliospheric transport model}\label{methods}

We have developed an electron acceleration and transport model for the inner corona and heliosphere, starting at $1$ solar radius $R_\Sun$. It is assumed that the extent of the simulated coronal region beyond the solar surface is always negligible compared to 1.0~AU (maximum 40~Mm vs approx. 150,000~Mm, or 0.026\%).

\begin{figure*}[t!]
    \centering
    \includegraphics[width=0.9\textwidth]{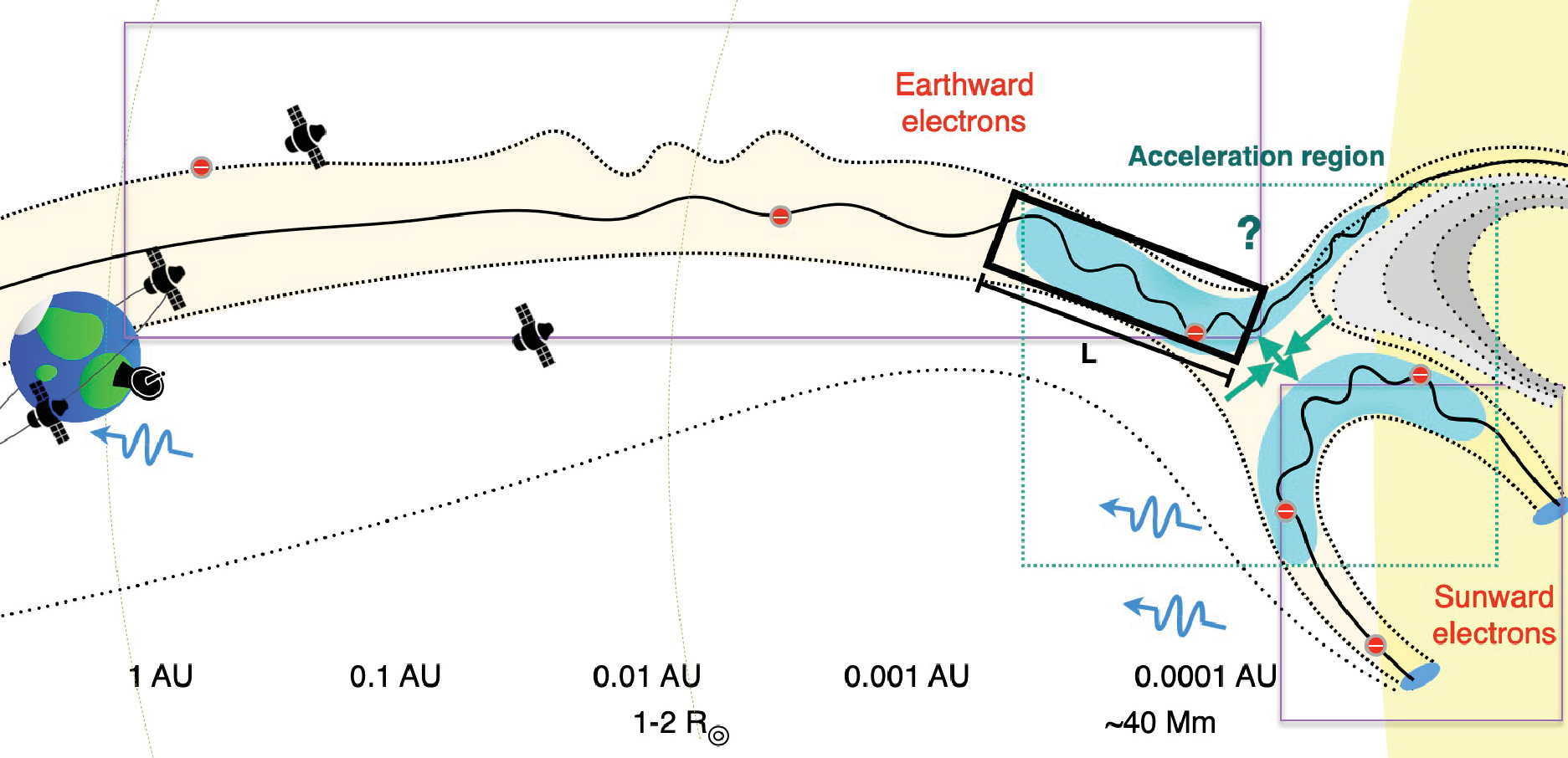}
    \caption{Simplified cartoon indicating one possible magnetic topology (such as emerging flux) that could lead to the production of both flare HXR-producing, and in-situ detected electrons in close proximity, in hot and over-dense regions close to the flare energy release site. In Paper I, we concentrate on studying the properties of those electrons accelerated during the flare but observed in-situ, by modelling the transport of an injected electron population through a hot and over-dense coronal region related to the flare (black-lined rectangle) of various temperature $T$, number density $n_{e}$ and size $L$, and then out into the heliosphere until 1.0 AU.}
    \label{fig:region_cartoon}
\end{figure*}

The model is composed of two discrete domains: a hot and over-dense (flare) coronal `acceleration' region of length $L$ with given uniform electron temperature $T$ and density $n_e$, and a sparser cold plasma ($T$ $\approx 0$) representing the wider heliosphere for $z > L$ (see Figure \ref{fig:region_cartoon}). Contrary to the name, the current study neglects acceleration mechanisms and will instead inject a single non-thermal electron distribution into a region dominated by collisional effects, henceforth referred to as the `collisional' region\footnote{The acceleration of the electrons out of the background thermal plasma will be discussed in Paper II.}. The extent and plasma properties of this collisional region are varied within reasonable limits to investigate the effect each variable has on the electron population ejected from this region. Following different solar flare observations of (above-the-) loop-top sources \citep[e.g.,][]{2014ApJ...781...43C,2015A&A...584A..89J,2020ApJ...900..192F}, we choose a sensible range of parameters for the coronal region: temperature $T$ ranging between $10-30$~MK, electron number density $n_e$ ranging between $10^{9}-10^{10}$ ~cm$^{-3}$ and region size $L$ ranging between $10-40$~Mm ($\sim14''-55''$).

The effects acting on electrons in the following transport equations are broadly separable into two types: collisional and non-collisional (and the latter into focusing and diffusion). In the heliospheric environment, adiabatic focusing and pitch-angle diffusion act in opposite to align or disperse particles (respectively) with regards to the magnetic field, the former being more significant where the magnetic field is strong and the latter where the mean free path is short. In the coronal region, collisional energy losses may dominate in the dense environment, whereas collisions will be negligible in the sparser heliospheric domain where non-collisional effects have previously been demonstrated to take precedence (due to low heliospheric densities described by empirical models and observations e.g., \citealp{newkirk1967structure,1977SoPh...55..121S,1998SoPh..183..165L,1999SSRv...87..185F,2018SoPh..293..132M}).

Forces due to gravity and electric fields are neglected. In this preliminary study, we neglect the effects of Langmuir wave turbulence and Landau damping that can produce a clear spectral break at $\sim40$~keV or lower (\citet{2009ApJ...695L.140K}; Section \ref{discussion} briefly discusses how wave-particle interactions may change the diagnostics outlined in Section \ref{results}). The effects of cross field particle diffusion and field line meandering which are vital for large spread events \citep{2016A&A...591A..18L} are neglected here.Particle (re-)acceleration in the heliosphere, such as that induced by shocks associated with co-rotating interaction regions (CIRs) \citep[as described by][]{Heber_1999,Allen_2021,Zhao_2019} and the heliospheric current sheet \citep[e.g.,][]{Zharkova_2012,Khabarova_2015} are also ignored in the present study.

\subsection{Governing Fokker-Planck Equation}

To describe the transport of a chosen electron distribution function $f(t,z,v,\mu)$ in time $t$, along a guiding magnetic field $z$, speed $v$ and cosine of the pitch-angle ($\beta$) to the guiding magnetic field $\mu=\cos\beta$, we use the following Fokker-Planck equation \citep[e.g.,][]{1981phki.book.....L,1986CoPhR...4..183K},

\begin{equation}\label{fp_eqn}
\begin{split}
\frac{\partial f}{\partial t}+ \mu v\frac{\partial f}{\partial z} = &\underbrace{- \frac{v(1-\mu^{2})}{2L_{z}}\frac{\partial f}{\partial \mu}}_\text{adiabatic focusing}\\
&\underbrace{+\frac{\partial}{\partial \mu}\left[D_{\mu\mu}\frac{\partial f}{\partial \mu}\right]}_\text{non-collisional pitch-angle scattering}\\
& \underbrace{+ \frac{\Gamma}{2v^{2}} \left[\frac{\partial}{\partial v}\left(2 v G(u) \frac{\partial f}{\partial v}
+4 u^{2} G(u) f\right)\right]}_\text{collisional energy losses}\\
& \underbrace{+ \frac{\Gamma}{2v^{3}} \left[ \frac{\partial}{\partial \mu}\left( (1-\mu^2) \biggl [ {\rm erf}(u) -G(u) \biggr ] \, \frac{\partial f}{\partial \mu} \right) \right]}_\text{collisional pitch-angle scattering}\\
\end{split}
\end{equation}

The first two terms on the right hand side of Equation \ref{fp_eqn} model a simple heliospheric environment with the presence of only adiabatic focusing and (parallel) pitch-angle scattering following e.g., \citet{1969lhea.conf..111R,2009ApJ...693...69D,2013JSWSC...3A..10A}. Both terms are present within the defined collisional coronal `acceleration' region $z<L$ and the wider heliosphere. $D_{\mu\mu}$ is the pitch-angle diffusion coefficient which quantifies the diffusion of an electron subject to its pitch-angle, velocity and local mean free path $\lambda$. The diffusion coefficient and its $\mu$ derivative are given by:

\begin{equation}
    D_{\mu\mu} = K (1 - \mu^2) (|\mu|^{q-1} + h)
\end{equation}
\begin{equation}
\frac{\partial D_{\mu\mu}}{\partial \mu}= K\mu\bigg[(q-1)(1-\mu^{2})|\mu|^{q-3} - 2(|\mu|^{q-1} + h)\bigg]
\end{equation}
where
\begin{equation}
    K = \frac{3 v}{2 \lambda (4-q) (2-q)}
\end{equation}
and $q=5/3$ is the spectral index of magnetic field fluctuations, taken as a Kolmogorov spectrum and constant $h=0.01$, which accounts for unmodelled scattering affects that are otherwise dominated at $|\mu| > 0$ in the heliosphere.
Following \citet{2018PhDT.......143A}, the mean free path $\lambda$ is calculated as in Equation \ref{lambda}, 

\begin{equation}\label{lambda}
    \lambda = \lambda_\Earth \Bigg(\frac{z}{z_\Earth}\Bigg)^\kappa \Bigg(\frac{p}{p_{\text{min}}}\Bigg)^{2 \xi}   
\end{equation}

dependent on the mean free path at 1.0 AU, $\lambda_\Earth = 0.3$ AU, the ratio of electron position $z$ relative to 1.0 AU ($= z_\Earth$) and the ratio of current and minimum electron momenta $p$ (derived from the minimum allowed kinetic energy in the heliosphere). $\kappa$ and $\xi$ are parameters that quantify the degree to which the electron momentum and radial distance from the Sun affect the mean free path. For this study they are given as $\kappa = 0.5$ and $\xi = -0.2$, and not explored further. The magnetic field at any point along $z$ is given by Equation \ref{B_field} \citep{1978SoPh...57..279D}, 

\begin{equation}\label{B_field}
    B = \frac{1}{2}\Bigg(\frac{z}{R_\Sun}-1\Bigg)^{-\frac{3}{2}}
\end{equation}
and $L_z$ in Equation \ref{fp_eqn} is ratio of the local magnetic field to its spatial gradient.
\begin{equation}
    L_z = \frac{B(z)}{\left(-dB/dz\right)}
\end{equation}

The final two terms in Equation \ref{fp_eqn} describe collisional energy losses and pitch-angle scattering respectively \citep[e.g., ][]{2014ApJ...787...86J,2015ApJ...809...35K} where $\Gamma = 4 \pi e^4$ln$\Lambda n_e / m_e^2$, for electron charge $e$ [statC], Coulomb logarithm ln$\Lambda$ and electron mass $m_e$ [g]. The error function erf$(u)$ and the Chandrasekhar function $G(u)$ are given by
\begin{equation}
    \text{erf}(u) \equiv (2 /\sqrt{\pi}) \int_0^u \text{exp}(-u^2) du
\end{equation}
and 
\begin{equation}
    G(u) \equiv \frac{\text{erf}(u) - u \text{erf}'(u)}{2u^2}
\end{equation}
where $u$ is the dimensionless velocity $u = v/(\sqrt{2} v_{th})$ and $v_{th} = \sqrt{k_BT_e/m_e}$. The error function and $G(u)$ control the lower-energy ($E\approx k_{B}T_e$) electron collisional interactions ensuring that they become indistinguishable from the background thermal plasma.

\subsection{Conversion to Stochastic Differential Equations (SDEs)}
The Fokker-Planck equation (Equation \ref{fp_eqn}) can be rewritten as a Kolmogorov forward equation \citep{Kolmogorov1931} and then converted to a set of time-dependent stochastic differential equations (e.g., \citet{1986ApOpt..25.3145G}; \citet{2017SSRv..212..151S}) that describe the evolution of $z$, $E$, and $\mu$ in It\^o calculus. The electron transport equations are solved numerically by a series of first-order Euler expressions, returning the evolution of electron velocity $v$, pitch-angle $\mu$ and position $z$ along the guiding field:

\begin{equation}\label{sdes}
    \begin{split}
    v_{i+1} & = v_{i} - \frac{\Gamma}{v_{i}^{2}} \left(\text{erf}(u_{i}) - 2u_{i} \text{erf}'(u_{i}) + G(u_{i})\right) \,\Delta t \\
    &+ \sqrt{\frac{2\Gamma G(u_{i}) }{v_{i}}\Delta t} \;W_v(t)
        \end{split}
\end{equation}
\begin{equation}
\begin{split}
    \mu_{i+1} & = \mu_{i} \\
    & - \frac{\Gamma \mu_{i} (\text{erf}(u_{i}) - G(u_{i}))}{v^3_{i}}\,\Delta t + \Bigg(\frac{d D_{\mu\mu}}{d \mu} + \frac{v_{i}(1 - \mu^2_{i})}{2 L_z}\Bigg)\,\Delta t \\ 
    & + \sqrt{\left(2 D_{\mu\mu} + \frac{\Gamma(1 - \mu_{i}^2)(\text{erf}(u_{i}) - G(u_{i}))}{v^3_{i}}\right) \,\Delta t} \;W_\mu(t)
    \end{split}
\end{equation}

\begin{equation}
    z_{i+1} = z_{i} + \mu_{i} v_{i} \Delta t
\end{equation}

$W$ is a value randomly chosen each time step from a Brownian normal distribution with a mean of 0 and variance of 1. This provides the stochastic element to the electron transport. In the coronal region, we set the time step $\Delta t$ at a constant value of $10^{-2}$~s; in order for collisional effects to have significance, there must be multiple simulation steps within the relatively small coronal region. With initial kinetic energies of order 10~keV, beamed electrons are expected to exit the largest modelled coronal region (40~Mm) within a few tenths of a second. Setting $\Delta t$ = $10^{-2}$~s allows even the most energetic electrons to have multiple simulation steps within the coronal region. 

As described in \citet{2009JCoPh.228.1391L} and \citet{2014ApJ...787...86J}, at velocities less than
\begin{equation}
v\leq\bigg (\Gamma\frac{8}{3} \sqrt{\frac{m_{e}}{2 \pi k_{B}T_e}}\Delta t \bigg)^{1/2}
\end{equation}
the analytical equation
\begin{equation}\label{eq: low_v}
v\simeq \bigg (v_{0}^{2}+\Gamma\frac{8}{3} \sqrt{\frac{m_{e}}{2 \pi k_{B}T_e}}\Delta t \bigg)^{1/2}
\end{equation}
can be used to determine $v$ to remove any divergence at low $v$ in Equations \ref{fp_eqn} and \ref{sdes}. For such velocities, $\mu$ can be drawn randomly from an isotropic distribution between -1 and +1.

Outside of the coronal collisional region, we use a time step $\Delta t$ = $1.0$~s and stop updating the velocity terms, as the mean free path in the simulated heliosphere is large enough that further collisional effects are assumed to be negligible enough to ignore. Any electrons with kinetic energies below 1~keV are taken to be thermal and have their pitch-angle values frozen, rather than randomised every step, saving computation time otherwise spent propagating electrons below energies of interest.

\subsection{Initial, plasma and boundary conditions}
At a singular point called $z=R_\Sun$, $10^6$ electrons are initialised with a beamed pitch-angle $\mu_0 = 0.99$, where positive values correspond to a direction outwards from the Sun and $\mu_0 = \pm 1.0$ is perpendicular to the solar surface. Since electrons are not accelerated from a thermal population in this initial study, we create a power-law velocity (energy) distribution between the energies of 5~keV and 100~keV, matching the electron energies detected via remote-sensing HXR observations at the Sun, with a power index in energy $E$ of $\delta=3$ ($E^{-\delta}$).

The electrons are initialised at the lowest point of a coronal region, simulated for a length of time sufficient for all particles to be ejected into the heliosphere. Heliospheric electrons are then simulated for an additional period of $t=50,000$s, after which the majority of the initial distribution have reached positions $z>1.0$ AU. This procedure is performed for all combination of coronal region properties.

As mentioned, we test the following coronal `flaring' conditions: temperature $T$ equal to either 10~MK, 20~MK or 30~MK, electron number density $n_e$ equal to $1\times10^{9}$~cm$^{-3}$, $5\times10^{9}$~cm$^{-3}$ or $1\times10^{10}$~cm$^{-3}$, and region length $L$ equal to 10~Mm, 20~Mm, 30~Mm or 40~Mm.

The kinetic energies and pitch-angles of the population for each set of coronal region properties are examined at various points in space and time. In particular, we will examine the time-integrated electron fluence (spectra) at one location in space, and the flux peak spectra, where the energy spectra are created using the peak flux in each energy bin or channel, at the following heliospheric locations of 0.01~AU, 0.4~AU and 1.0~AU. It is assumed that electrons are transported out to empty space, such that no planetary magnetospheres are present in the simulation.

As the transport equations do not inherently prevent an unphysical pitch-angle value ($|\mu|>1$), care must be taken when modelling particles close to this boundary. In this case they are reset in the same position with a uniformly random pitch angle $|\mu| = [0.89,1.00)$. Particles cannot be transported below $R_\Sun$ and are perfectly reflected at this boundary. The collisional region is not present in the heliospheric component of the simulation, and so any ejected electrons that would re-enter those ranges of distance from the Sun are similarly reflected. This does not represent a physical effect, but rather a way to prevent extended thermalisation of particles that may be trapped close to the boundary without affecting statistics. We find that in all cases the coronal ejecta population remains relatively anisotropic, such that electrons that do re-encounter the boundary represent a very small proportion of the heliospheric energy spectra.

\section{Results}\label{results}

\begin{figure*}[hpbt!]
    \centering
    \includegraphics[width=0.9\textwidth]{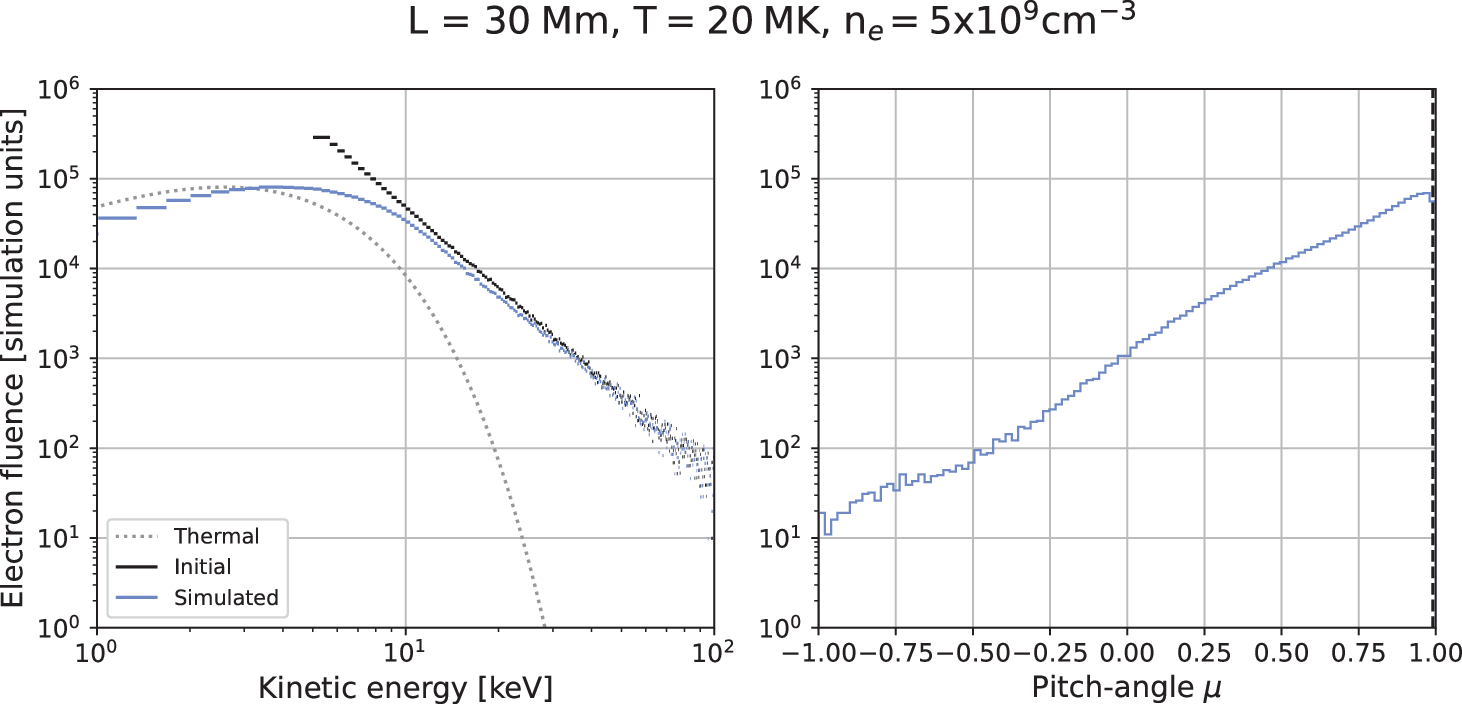}
    \caption{Example ejected electron fluence (blue) versus energy (left) and pitch-angle (right). The injected distribution (power law, beamed) is denoted by the black curve in the left plot and by the dashed black line in the right. The injected distribution travels through a `flare' region with the following plasma properties: length $L=30$ Mm, temperature $T=20$~MK and number density $n_e=5\times10^{9}$~cm$^{-3}$. A thermal distribution corresponding to a plasma temperature of $T=20$~MK is also shown (grey dotted, left)}
    \label{fig:cor_fluence_example}
\end{figure*}

\subsection{Collisional region plasma properties}

The ejected electron energy and pitch-angle fluence distributions of an example parameter set are shown in Figure \ref{fig:cor_fluence_example}. The ejected fluence corresponds to time-integrated distributions measured at one spatial location (similar to spacecraft data), here namely the furthest boundary from the Sun of the simulated collisional region of length $L$ ($z = R_\Sun + L$). The ejecta fluence distribution (blue) is compared with the injected beamed power law population (black) and a thermal distribution corresponding to the ambient plasma temperature (grey dotted) of $T=20$~MK. Prior to ejection, the population steadily evolves towards a thermal shape, though the high-energy tail ($>$10~keV) remains relatively unchanged from injection, as expected in lower density ($n_e=5\times10^{9}$~cm$^{-3}$) plasma. These electrons are ejected quickly and do not undergo enough collisions to begin to significantly thermalise above 10~keV. The population as a whole remains relatively anisotropic out to the point of ejection, though some pitch-angle scattering is evident if largely insignificant\footnote{Here, there is an obvious bias towards high positive pitch-angles since we inject a beamed distribution, and measurement is taken at the first crossing of a given boundary. However, the aim here is to study the effects of different plasma properties on a given distribution, not exhaustively examine different pitch-angle injections.}. We examine these properties in the respective fluence distributions of all parameter sets in order to determine how they are affected by changing the region size, plasma temperature and density. 

\begin{figure*}[hpbt!]
    \centering
    \includegraphics[width=0.9\textwidth]{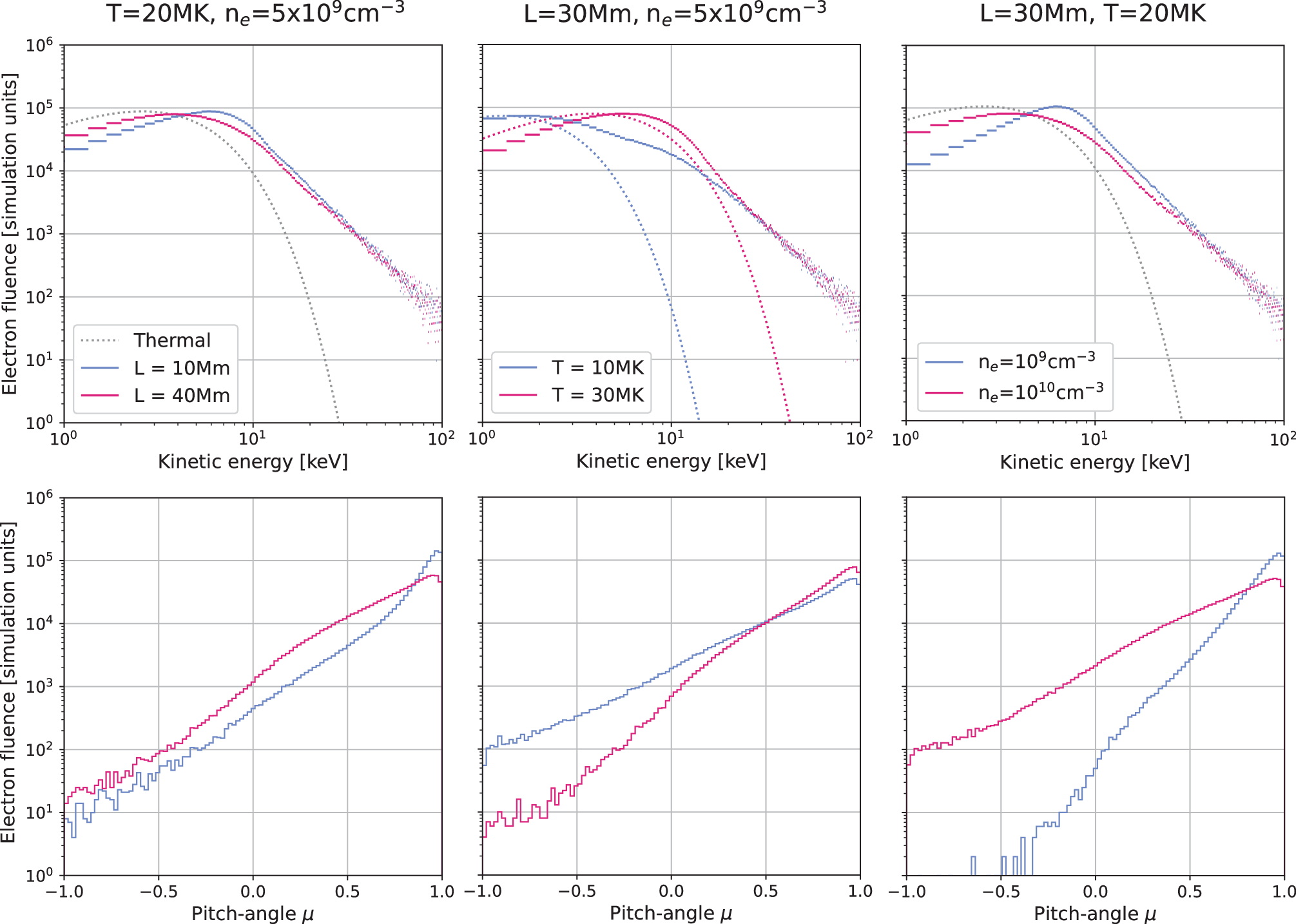}
    \caption{Collisional-boundary electron energy (upper) and pitch-angle (lower) fluence distributions independently varying each parameter between their minimum (blue) and maximum (pink) values, while fixing the other parameters at intermediate values. Dotted lines represent the thermal distributions associated with the given temperature values. The fluence peaks at higher energies for smaller, hotter and sparser regions, which also correspond to less isotropic ejecta distributions. As expected, the electron spectra above 20 keV remains relatively unchanged regardless of region parameters.}
    \label{fig:cor_plasmaparams_compare}
\end{figure*}

In Figure \ref{fig:cor_plasmaparams_compare} these distributions are compared for the minimum and maximum tested values of $L$, $T$ and $n_e$ in turn, fixing the other two parameters at intermediate values. In all parameter sets, the fluence (energy) distribution remains more or less unchanged above 20~keV, as these electrons are so energetic as to be practically unaffected by the plasma prior to ejection for the properties we test for. As such the focus of this analysis will be on energies below this approximate value, in particular in the 1-20~keV range, where spectral variations due to hot, over-dense plasma environments should show, if present.

Changing the physical size of the region shows the smallest change in the energy and pitch-angle distributions. The smaller region has a fluence peak at a slightly higher energy and fewer electrons ejecting at lower energies. Smaller regions also appear to have somewhat less isotropic ejecta. As the electrons are travelling across a shorter distance (and so a shorter time) before ejection, the population has had less time to thermalise than for a larger region, therefore resulting in the distributions retaining more of the features of the non-thermal injected population.

Between plasma temperatures of 10~MK (blue) and 30~MK (pink), a higher $T$ shows a higher energy for the fluence peak. In similarly sized regions or equivalent density, hotter ambient temperatures result in the population being thermalised to higher energies before ejection. Such temperatures also appear to make the population less isotropic than for colder regions (since the collisional times are increased in a hotter plasma).

In higher density regions, the population evolves more towards a thermal distribution before ejection than for a lower density, becoming more isotropic in the process. This is due to the greater number of collisions an average electron will undergo in this time, making the process of thermalisation more efficient. Conversely, the population in the lower density region has fewer average collisions and retains more of the injected distribution's shape.

\begin{figure*}[hpbt!]
    \centering
    \includegraphics[width=0.9\textwidth]{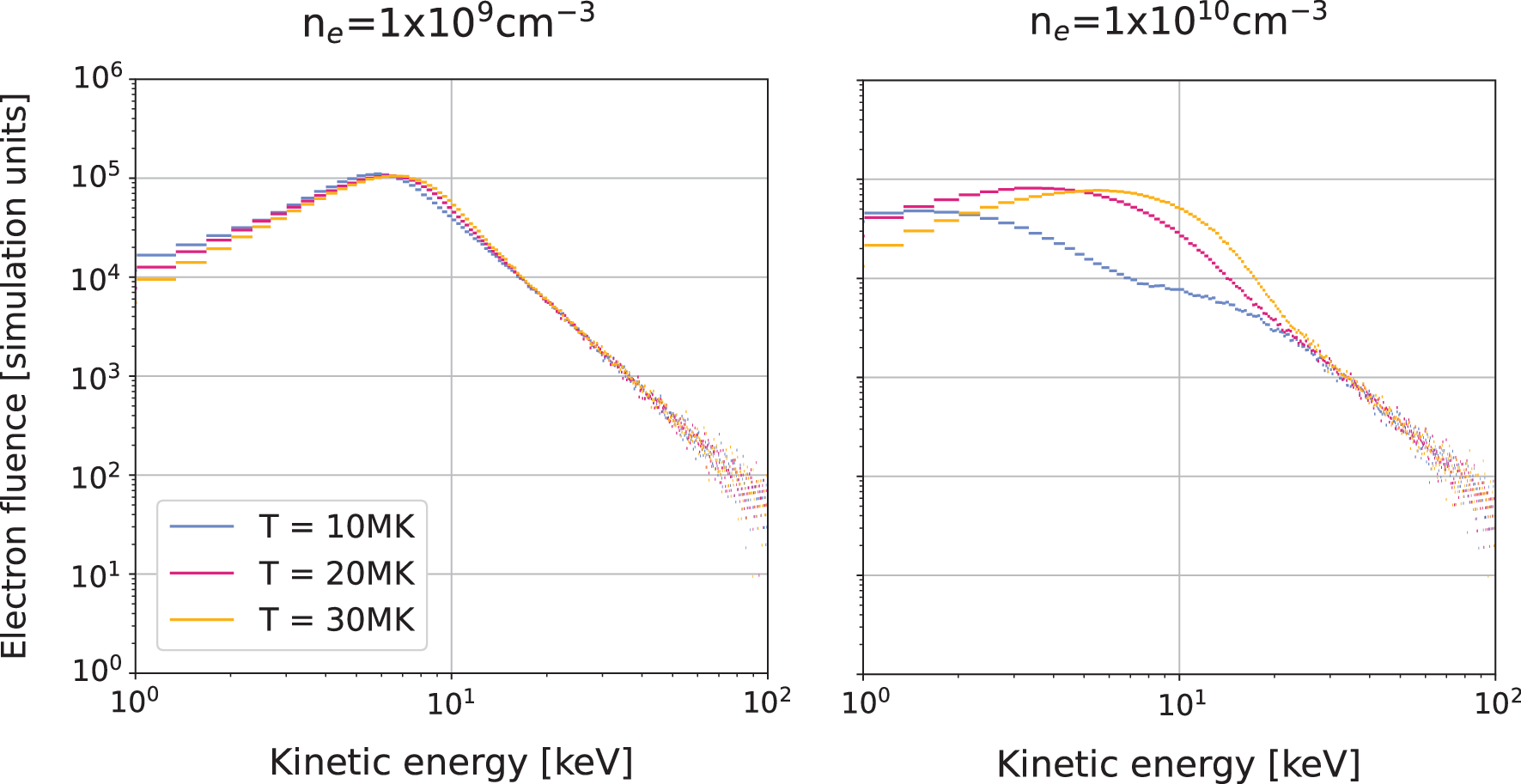}
    \caption{The temperature dependence of the ejected electron fluence distributions is compared for minimum (left) and maximum (right) simulated densities, and $L=30$~Mm. Distribution shape is very similar in all low-density regions, but the spectral differences due to temperature can be clearly seen in higher density cases.}
    \label{fig:cor_fluence_nT}
\end{figure*}

The effects of plasma temperature and density are not entirely separable, as both play a role in the electron collisional time and by extension the efficiency of thermalisation. In Figure \ref{fig:cor_fluence_nT} we compare how varying the electron temperature for high and low densities affects the resulting fluence spectra at the collisional region boundary.

In the low-density case, there are too few collisions for the distribution to thermalise above the lowest measured energies before ejection. The fluence peaks have approximately the same heights and correspond to the same band of kinetic energy. At higher densities however we see a much more significant divergence between the fluence distributions. Low temperatures greatly suppress the fluence peak and decelerate the injected distribution to energies of order $\sim1$ keV. For higher temperatures, the fluence peaks have comparable heights to the low density region but show more electrons being thermalised between $1-10$ keV energies, corresponding to the thermal energy of the ambient plasma. It can be seen that ejecta from low-density regions have little temperature dependence making it more difficult to constrain the region properties based on ejecta distributions than it would be for a higher density. Differences in lower ($10$~MK) temperature and higher temperature ($\ge20$~MK) can be clearly distinguished in high density plasma ($10^{10}$ cm$^{-3}$). However, for any temperature plasma, there are clear spectral differences between the low and high density cases.

\begin{figure*}[t!]
    \centering
    \includegraphics[width=0.7\textwidth]{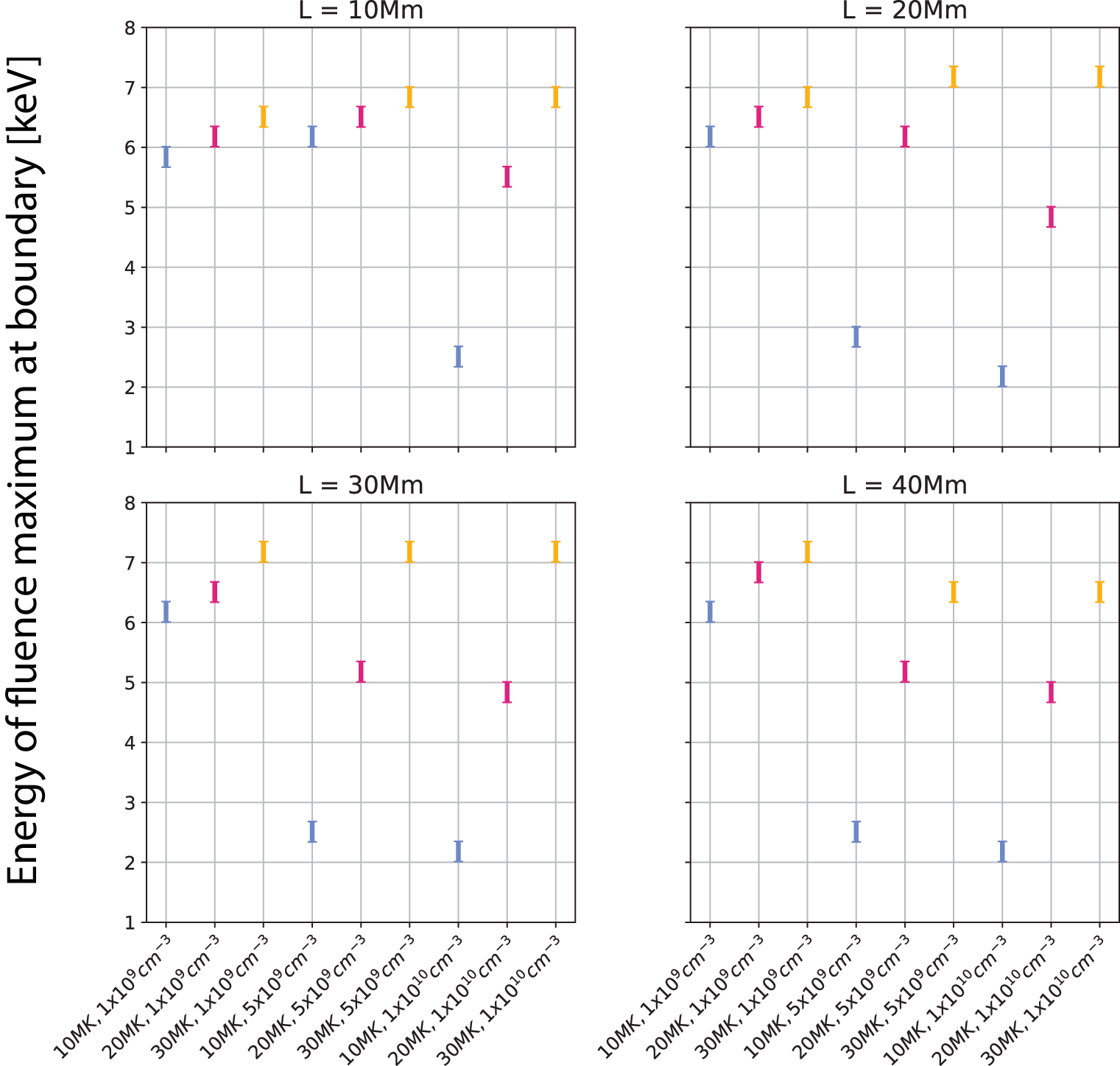}
    \caption{Peak (in energy) of ejected electron fluence for every simulated parameter set, with each window corresponding to a different region size. The parameter set labels (x-axis) are formatted in order of region temperature and electron density. Increasing the electron density of the region significantly reduces the energy at which the fluence peaks for most region sizes, with a greater degree of deceleration for colder plasmas.}
    \label{fig:cor_peaks}
\end{figure*}

In Figure \ref{fig:cor_peaks}, for the ejected fluence spectra at the collisional region boundary, we plot the energy at which the fluence spectra peaks for each studied plasma region varying in $L$, $T$ and $n_e$. We determine if such properties are useful to help diagnose the plasma properties of the flare `acceleration' region. At higher densities the energy at which the fluence peaks drops significantly at all temperatures, though to a lesser degree in hotter regions. Higher temperatures themselves increase this peak energy in regions of equal density, with seemingly diminishing returns at $L \ge 20$ Mm and $n_e \ge 5\times10^9$ cm$^{-3}$.

For each $L$, we see the general pattern of energy of the fluence peak increasing with hotter plasma temperature and decreasing with increasing plasma number density (with changes due to temperature more evident in higher density regions). In this example, a peak fluence sitting at an energy of $\sim5$~keV could correspond to $T=20$~MK (high flare temperature), with $L$ ranging between $20-40$~Mm and $n$ between $5\times10^9-1\times10^{10}$~cm$^{-3}$ (i.e., higher density). We will now extend the use of these diagnostics into the heliosphere where such spectra can be observed.

\subsection{Extending transport to the heliosphere}

The distributions for energy, pitch angle, and ejection time generated by the coronal component are taken as injection profiles and iterated out to 1.0 AU. In order to draw an approximate equivalence with the energy resolution of modern spacecraft in the heliosphere (and archived data), all data discussed after this point will be displayed with 1 and 3 keV binning unless otherwise stated.

\begin{figure*}[hpbt!]
    \centering
    \includegraphics[width=0.9\textwidth]{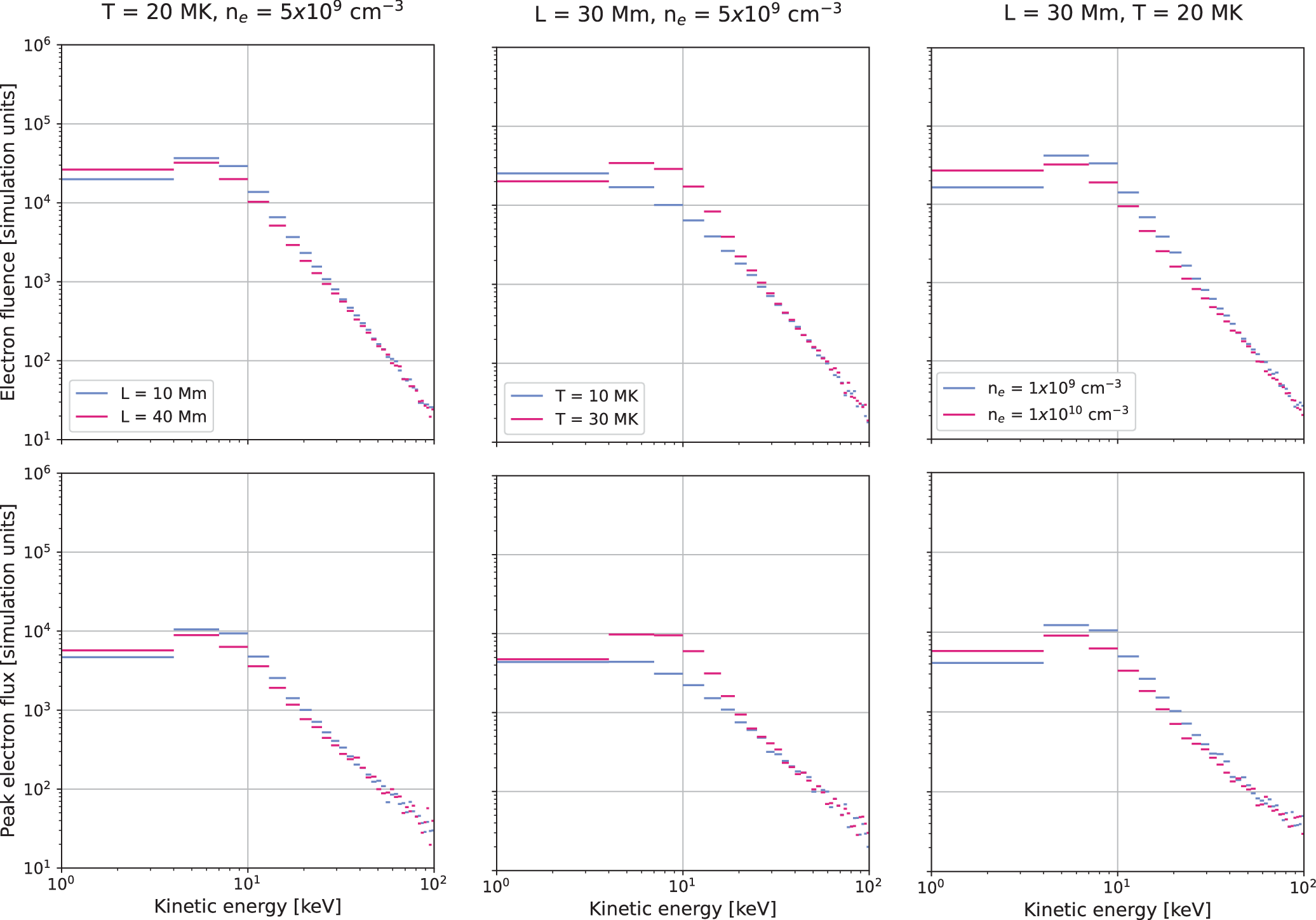}
    \caption{Fluence (upper) and peak flux (lower) distributions of the ejected population at 1.0 AU with a simulated energy binning of 3~keV for the minimum and maximum collisional region parameters (shown in each legend). The fluence and flux distributions peak at higher energies for hotter and sparser collisional regions similar to the region boundary results, with the largest divergence found between minimum and maximum temperature parameter sets.}
    \label{fig:hel_plasmaparams_compare}
\end{figure*}

Figure \ref{fig:hel_plasmaparams_compare} shows the fluence and peak flux energy spectra at 1.0 AU for independently varied values of $L$, $T$ and $n_e$ (similar to Figure \ref{fig:cor_plasmaparams_compare}). There are minor differences between both distributions for a small and large collisional region, suggesting that region size is difficult to constrain within the selected range.

Here, the most significant difference is between colder and hotter regions, with the maximum fluence and peak flux occurring at larger energies as the temperature increases, as expected. It also shows a significant deviation between the low and high temperature regions in the 10-19~keV range that is not as readily seen when other parameters are varied instead. For 20~MK, the density comparison demonstrates a similar (if weaker) pattern, with lower density regions allowing more of the non-thermal injected population to escape into (and propagate through) the heliosphere.

\begin{figure*}[hpbt!]
    \centering
    \includegraphics[width=0.9\textwidth]{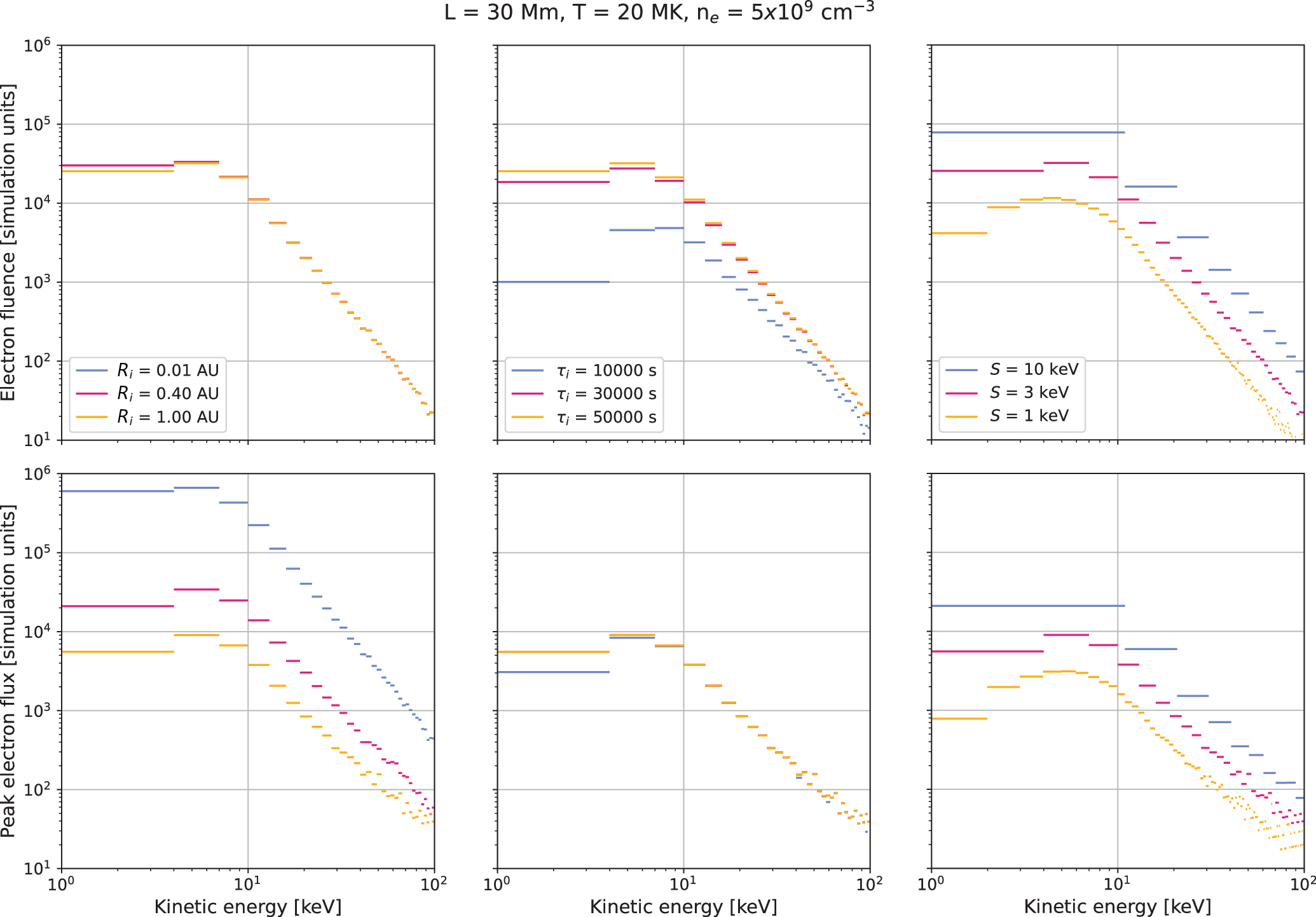}
    \caption{Fluence (upper) and peak-electron flux (lower) spectra of the ejected population for intermediate collisional region plasma properties and varying analytical parameters: the distance of the virtual measurement from the Sun $R_i$ (left), the total integration time of the simulation $\tau_i$ (middle) and the energy bin resolution $S$ (right). Other analytical parameters are fixed at  $R_i = 1.0$~AU, $\tau_i = 50,000$~s and $S = 3$~keV. The fluence is effectively unchanged across the heliosphere at the maximum integration time with greater peak flux values at all energies closer to the Sun. Shorter integration times show a significant reduction of fluence at 1.0~AU. Bin sizes larger than $\sim 3$ keV mask the maximum values of peak flux and fluence and make it difficult to identify features at the low-energy ranges ($<$ 10 keV) of both distributions.}
    \label{fig:hel_analysisparams_compare}
\end{figure*}

In addition to varying the physical parameters of the collisional region, we can also vary the parameters of the analysis itself: accounting for the position of a virtual instrument $R_i$, the total simulated runtime $\tau_i$ and the width of the energy bins $S$. These are displayed in Figure \ref{fig:hel_analysisparams_compare}, with fixed values of $R_i = 1.0$~AU, $\tau_i = 50,000$~s and $S = 3$~keV unless otherwise stated.

For the maximum simulated runtime, the fluence across the heliosphere remains relatively unchanged as the majority of electrons will cross all values of $R_i$ that are being compared. The peak flux on the other hand is significantly higher closer to the Sun, though the shape of the distribution remains approximately the same. Scattering effects in the heliosphere delay transport along a 1D line out to 1.0 AU so while the energy values are unchanged and will arrive at every point along the heliosphere eventually, the rate of arrival is spread out resulting in lower peak flux. 

Reducing the simulated runtime shows a lower number of electrons arriving at 1.0 AU, reducing the overall fluence. The lowest energy electrons are expected to arrive latest so reducing $\tau_i$ steadily will show a much faster decrease in low energy fluence first. This is also reflected in the flux peak plot though to a lesser extent.

Enlarging the bin size results in a higher fluence and peak flux per bin, as naturally more energy is collected within a wider range. However this has the added effect of obfuscating the location of the spectral peaks. Sufficiently high resolution energy measurement would be required to identify and measure this feature ($\sim<3$~keV binning at energies $<20$~keV).

\begin{figure*}[hpbt!]
    \centering
    \includegraphics[width=0.84\textwidth]{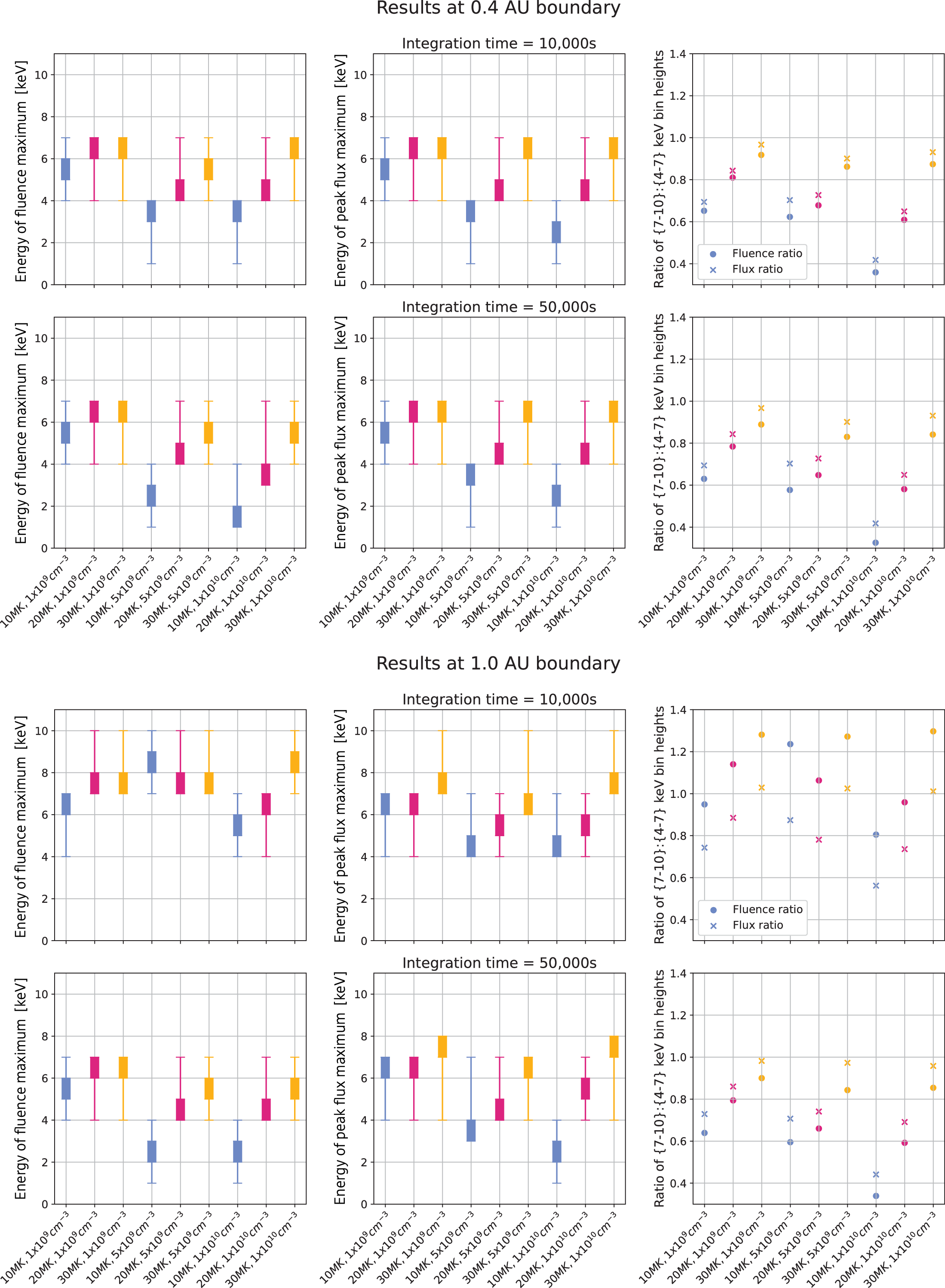}
    \caption{Full collisional-region parameter analysis at 0.4 AU (upper) and 1.0 AU (lower). The energies where the maximum fluence (left) and peak flux (right) occur are shown for 3~keV (thin line) and 1~keV (thick line) resolutions. Ratios of the [$7-10$) to [$4-7$) keV bin heights for both fluence (circles) and flux (crosses) are calculated and displayed in the right column. While the size of some 1 and 3 keV bins mask the change in peaks (fluence and peak flux) between parameter sets for regions with temperatures above 10 MK, the bin height ratios demonstrate the same relation for a sufficiently long integration time.}
    \label{fig:hel_peaks}
\end{figure*}

Compiled in Figure \ref{fig:hel_peaks} are the energies where the maximum values of fluence and peak flux occur, for electron distributions at two heliospheric positions (0.4~AU and 1.0~AU) and integration times (10,000~s and 50,000~s). We also consider how they would manifest in 1 and 3 keV binning schemes comparable to modern spacecraft instrumentation. In addition, we calculate the ratio of the [$7-10$) to [$4-7$) keV bin heights for each parameter set. With the exception of the $R_i = 1.0$~AU, $\tau_i = 10,000$~s case, the peaks and ratios show relatively `settled' cases where the majority of electrons have reached $R_i$ and little variation can be seen between these parameter sets. At high $R_i$ and low $\tau_i$ the maxima (fluences and peak fluxes) are biased towards higher energies that have been able to reach 1.0 AU in the reduced integration time. In addition, this is the only case where the fluence bin ratios exceed those of the flux ratios, indicating that the peak flux ratios are identifiable within a shorter simulation runtime than the fluence ratios.

With regards to the collisional region parameters, there are clear patterns shown in both the peaks and ratios. For the high-resolution 1-keV-binned peak cases, it can be seen that increasing the temperature of the region results in a fluence and flux peaks occurring at a higher energy at both 0.4 and 1.0 AU, though there is relatively little difference when increasing the density at high temperatures. 

Increasing the density at lower temperatures on the other hand shows a significant drop in peak fluence and flux energies from $1\times10^9$ cm$^{-3}$ to $5\times10^9$ cm$^{-3}$. This indicates that a cold ($<$ 30 MK) and dense ($>$ 1x10$^9$ cm$^{-3}$) collisional region is an effective decelerator of energetic electrons and evident even out at 1.0 AU.

These patterns are somewhat more difficult to discern at 3~keV binning; with the exception of the lowest temperature parameter sets, the high-resolution peaks occupy the same [$4-7$) keV bin. Instead we compare the relative heights of the [$7-10$) and [$4-7$) keV bins for fluence and peak flux, which show the same relations as the higher-resolution maxima. As such, in the absence of sufficiently high-resolution energy binning these ratios are our preferred metric for constraining collisional region properties using heliospheric data.

\section{Summary and Discussion}\label{discussion}

Locating the acceleration region is fundamental in understanding how and where energetic particles are produced in flares, as well as the relationship between different particle populations observed at the Sun and in interplanetary space (e.g., plasma environment, magnetic topology). Thus, examining whether signatures of hot, over-dense plasma exist in interplanetary electron spectra is vital. This analysis is concerned with so-called `prompt' events where energetic electrons are evidenced to be directly related to the flare and not to any secondary acceleration mechanism such as a CME shock. Previous studies provide conflicting evidence regarding where in-situ electrons are energised compared to their HXR-emitting counterparts. In certain flares, for example, emerging flux or interchange reconnection \citep[e.g.,][]{2023A&A...670A..56B, 2016SoPh..291.1357W, 1977ApJ...216..123H} and/or the presence of a flare-jet structure \citep[e.g.,][]{2020ApJ...889..183M,2011ApJ...742...82K,2009A&A...508.1443B} may ultimately allow loop-top electrons to escape, while having access to hot, over-dense material usually related to HXR-emitting electrons. Alternatively, or in conjunction, the presence of a turbulent acceleration mechanism \citep[e.g.,][]{2021ApJ...923...40S, 2017PhRvL.118o5101K} may both heat the surrounding plasma and energise electrons simultaneously.

Here, we analysed how the plasma properties of a hot, over-dense coronal region (including or close to an acceleration region) change the properties of $\approx1-20$ keV electrons and examine whether the signature of such plasma properties can be extracted from the electron fluence or peak flux spectrum. In this study, we only present very specific or `extreme' plasma examples that are easily identifiable in the electron spectrum, unlike lower temperatures ($<10$~MK) and densities ($<10^{9}$~cm$^{-3}$), possibly in higher altitude locations away from flare site producing HXR-emitting electrons. 

The results suggest that if interplanetary flare-related electrons are indeed produced in hot, over-dense regions, possibly during continued accelerated in flares (similar to HXR-producing electrons) and within a magnetic topology that allows the escape of such electrons, then we should be able to detect these signatures, and even constrain the plasma properties of such a region. We tested how different bin sizes used for plotting electron fluence or peak flux affected the extraction of such properties, and how electron loss at different locations in the heliosphere can also hide these signatures. Thus, the lack of such ‘thermal’ signatures would suggest that such particles indeed originate from ‘higher’ coronal locations, unrelated to the acceleration region producing HXR-emitting electrons. In order to diagnose these properties, a detailed individual flare study would have to be performed (including a comparison of in-situ and HXR-emitting electrons), beyond the scope of this initial modelling study.

Properties of an `acceleration' region can be more narrowly constrained by identifying signatures of the `thermal' ($<20$ keV) electron datasets. Energy of peak fluence and peak flux bands at resolutions of 1 keV both vary significantly between different parameters sets. If these features exist, then these properties can be readily extracted from modern in-situ instruments at comparable resolutions, meaning that such constraining is currently possible with existing and upcoming observational data. Such information will be extracted by carefully examining individual flares spectra with simulation outputs. The prospect of different plasma conditions changing the properties of an electron population (accelerated out of the thermal plasma) at higher energies (i.e., spectral index) will be studied in detail in Paper II.

The model we used in this preliminary analysis is rather artificial; a chosen accelerated electron distribution (single power law) is injected and transported through a hot over-dense region. Paper II will perform more realistic simulation where electrons are accelerated out of different thermal plasma using a turbulent diffusion model \citep[similar to][]{2023arXiv230113682S}, with the resulting shape of the spectra also dependent on the plasma properties of the acceleration region, and producing a smoother spectral transition from thermal to non-thermal if this region exists within the $1-20$~keV data. As mentioned, we purposely used higher temperatures and higher densities akin to hot flaring loops to show the types of spectra and diagnostics we may expect to see if such electrons are being produced in regions identical to HXR-emitting electrons. Moreover, we did not exhaustively input a range of electron spectral indices or low energy cutoffs into the simulation. We investigated the effect of independently decreasing the minimum initial energy (down to 3~keV) and increasing the spectral index of the injected distribution (up to $\delta=5$). The result of both changes was a slight decrease in the energy of peak fluence at 1.0 AU, by 0.5 to 0.7 keV. There is a similar drop in the [$7-10$):[$4-7$) keV bin ratios, by 0.1 to 0.2. These indicate that making such changes to the injected distribution will result in the fluence distribution at 1.0 AU shifting slightly towards lower energies. 

As discussed earlier, \cite{2009ApJ...695L.140K} investigated the transport of solar flare energetic electrons in the heliosphere taking into account the self-consistent generation and absorption of Langmuir waves, effects of nonuniform plasma, collisions, and Landau damping, and found that such processes lead to a spectral break and flattening below approximately 40~keV, acting similar to collisional effects in dense plasma, leading to an overall flatter spectra at lower energies. For our study, modelling is required to analyse the effects fully. However, we suggest that such processes should lead to flatter spectra in the $1-20$~keV range, with any visible peak due to a (partially-)thermal component possibly shifted to lower energies. 

In real data the flare-accelerated population must be separated from the different populations of the solar wind (core, halo and super-halo) \citep[e.g.,][]{1975JGR....80.4181F,2001Ap&SS.277..195P,2012ApJ...753L..23W}. Although flares are evidenced by their increased electron flux compared to the background, at $\approx 1$~keV the flare-spectra will start to merge back into the solar wind halo and core \citep[e.g.,][]{1984SoPh...91..345P,1985SoPh..100..537L,2022RvMPP...6...12W} and produce a spectral upturn close to merging point at $\approx 1$~keV. Thus as above, we suggest the $1-20$~keV range is the optimal range for such analysis. Upturns in the spectral data within the $1-20$~keV suggest the presence of a hotter thermal component. Interestingly, in a rare published example of electron peak intensity energy spectra i.e., \citep{2023arXiv230103650J} produced by the SolO Energetic Particle Detector (STEP, EPT and HET) \citep{2020A&A...642A...7R} we see a flattening i.e., \cite{2009ApJ...695L.140K} but then a noticeable upturn in the energy spectrum at lower energies (at around $10-20$~keV), which may indicate the signature of the accelerating plasma environment. Nevertheless, this provides an excellent example of why this spectral region deserves more attention and analysis.

\section*{acknowledgments}
RP \& NLSJ gratefully acknowledge the current financial support from the Science and Technology Facilities Council (STFC) Grant ST/V000764/1. The authors acknowledge IDL support provided by STFC. NLSJ is supported by an international team grant \href{https://teams.issibern.ch/solarflarexray/team/}{“Measuring Solar Flare HXR Directivity using Stereoscopic Observations with SolO/STIX and X-ray Instrumentation at Earth}” from the International Space Sciences Institute (ISSI) Bern, Switzerland. The data that support the findings of this study are available from the corresponding author upon reasonable request. We thank Professor Eduard Kontar for insightful comments.

\bibliography{main}
\bibliographystyle{aasjournal}

\end{document}